\documentclass[12pt,equations,fleqn]{article}
\usepackage{epsf}
\begin{document}
\newcommand{\half}{\frac{1}{2}}
\newcommand{\onlinecite}[1]{\cite{#1}}
\begin{titlepage}
\def\baselinestretch{1.4}
\null
\begin{center}
{\LARGE
First and second order transitions in dilute O($n$) models}
\vskip 10mm
{\Large Wenan Guo~$^{\S}$\footnote{present address: 
Physics Department, Beijing Normal
University, Beijing 100875, P. R. China}, 
Henk W.J. Bl\"{o}te~$^{\S \dag}$\footnote{e-mail: bloete@tn.tudelft.nl} \\
and Bernard  Nienhuis~$^{\ddag}$ \\}
\vskip 5mm
{\em
$^{\S}$ Faculty of Applied Science,  P.O. Box 5046,
2600 GA Delft, The Netherlands \\
$^{\dag}$ Instituut Lorentz,
Universiteit Leiden, Niels Bohrweg 2,
Postbus 9506, 2300 RA Leiden, The Netherlands\\
$^{\ddag}$ Instituut voor Theoretische Fysica,
Universiteit van Amsterdam, Valckenierstraat 65,
1018 XE Amsterdam, The Netherlands\\}
\end{center}
\vskip 5mm
\def\baselinestretch{1.6}
\begin{abstract}
We explore the phase diagram of an O($n$) model on the honeycomb lattice
with vacancies, using finite-size scaling and transfer-matrix methods.
We make use of the loop representation of the O($n$) model, so that
$n$ is not restricted to positive integers. For low activities of
the vacancies, we observe critical points of the known universality
class. At high activities the transition becomes first order. For $n=0$
the model includes an exactly known theta point, used to describe a
collapsing polymer in two dimensions. When we vary $n$ from 0 to 1, we
observe a tricritical point which interpolates between the universality
classes of the theta point and the Ising tricritical point. 
\end{abstract}
\vfill
\vskip 15mm
\end{titlepage}

\section{Introduction }
In two-dimensional models, exact solutions and Coulomb gas methods have
led to the discovery of a large collection of universality classes,
including some that depend on a continuously variable parameter (see
e.g. Refs.~\cite{Bax,3CG}), such as the Baxter model, the critical
and tricritical Potts models, and the critical O($n$) model.
However, such results are not known for the tricritical points of the
O($n$) model with general $n$, in spite of the fact that there are no
obvious reasons why such a family of tricritical points should be absent.
Tricritical behavior is known only for discrete values of $n$: the 
so-called theta point at $n=0$ \cite{DS} and the Blume-Capel tricritical
point \cite{Blume,Capel} at $n=1$, where the model intersects with the
$q=2$ Potts model, so that the results for Potts tricriticality apply.
The branch of $q$-state Potts critical points applies to a limited range 
of $q$ values. Its analytic continuation beyond the end point at $q=4$
appears to cover the same range of $q$, and is then found to describe a  
branch of tricritical points \cite{BNtrcrPotts}.
In the case of the O($n$) model on the honeycomb lattice, the situation
is similar in a sense: two branches of critical behavior are known
\cite{3Nhon}, which are the analytic continuation of each other, and
connect at the upper boundary at $n=2$. However, the analytic
continuation of the critical branch does not describe a phase
transition: instead, it describes the low-temperature O($n$) phase
which is still critical in the sense that the correlations decay
algebraically for general $n$.

Several other branches of critical behavior are known for O($n$)
models with additional interactions, see e.g. Ref.~\cite{3BN} for an
analysis of the square-lattice O($n$) model. One of these branches,
called branch 3, marks the boundary between a range of ordinary O($n$)
transitions and a first-order range, just as expected for tricritical
points. However, this higher critical point of branch 3 is accompanied
by the freezing out of Ising-like degrees of freedom \cite{3BN,GBN}
and does not match the known tricritical behavior at $n=0$ and 1.

Another family of critical points,
called branch 0, describes a dilute form of the O($n$) model. It
includes a point at $n=0$ that belongs to the universality class of
the theta point. In this sense branch 0 is a generalization of the
theta point to other values of $n$. Thus it could be a logical
candidate for a class of tricritical points, but the $n=1$ point of
branch 0 represents the F model \cite{Rys} and not the tricritical
Ising model.

This somewhat puzzling situation raises some questions, such as what
precisely is an O($n$) tricritical point, and what sort of connection
exists between the theta point and the Ising tricritical point?
As for the first question, we adhere to the accepted picture at $n=0$
and $n=1$ that a gradual increase of the activity of the vacancies will
eventually drive the O($n$) phase transition first order; the point
where the first-order line meets the line of continuous transitions
is considered a tricritical point. As for the second question, we
may look for a regime of tricritical points that interpolates between
a theta point and an Ising tricritical point.
In the absence of exact results we use a numerical method,
namely finite-size scaling of transfer-matrix results, in order to
explore such tricritical points for general $n$. 

Thus we select a model for our study that includes vacancies explicitly.
It includes the theta point defined by Duplantier and Saleur \cite{DS}
and the honeycomb O($n$) model without dilution \cite{3Nhon}
as special cases. For $n=1,2,\cdots$ this model is equivalent with
an O($n$) symmetric spin model \cite{3DMNS}.
The model and the numerical methods are described in Section 2. The 
results and their interpretation are presented in Section 3, and our
conclusions and outlook are summarized in Section 4.

\section{The model and numerical procedures }

We define a dilute O($n$) loop model in which the loops run on the
edges of the honeycomb lattice. Each edge covered by a loop has weight
$w$, and each loop has a weight $n$. Dilution is represented by means
of face variables: the elementary hexagons of the lattice are `vacant'
with weight $y$ and `occupied' with weight $1-y$. The loops are
forbidden to touch any vacant faces.

The occupied faces form a lattice ${\cal L}$ which is
any subset of the dual (triangular) lattice.
The loop configurations of this model are represented by graphs
${\cal G}$ consisting of non-intersecting polygons, which avoid the
edges of the vacant hexagons. Thus the partition function is
\begin{equation}
Z=\sum_{\cal L}\sum_{\cal G|L}y^{N_{v}}(1-y)^{N-N_{v}}w^{N_{w}}
n^{N_{l}}
\label{Zloop}
\end{equation}
where $N_{v}$ is the number of empty faces specified by the vacancy
configuration ${\cal L}$.  The number $N_{w}$ of edges covered by
${\cal G}$ is equal to the number of vertices visited by a polygon,
and $N_l$ the number of loops.  The second summation is only over
those graphs ${\cal G}$ allowed by ${\cal L}$. 

For the construction of the transfer matrix we choose the usual geometry
of a model wrapped on a cylinder, such that one of the lattice edge
directions runs parallel to the axis of the cylinder. The finite size
$L$ of the model is taken to be the number of hexagons spanning the
cylinder; the unit of length is thus $\sqrt{3}$ times the nearest-neighbor
distance. The transfer-matrix index represents the loop connectivity,
i.e. the way in which the dangling bonds at the end of the cylinder are 
mutually connected, and the vacancy distribution on the last row of 
the lattice. 
The introduction of vacancies leads to an increase of the number of 
possible values of the transfer-matrix index for a given system size.
Therefore
the present calculations are restricted to somewhat smaller system sizes
than e.g. those investigated in Ref.~\cite{GBN}.


Our numerical analysis focuses in particular on the asymptotic
long-distance behavior of the O($n$) spin-spin correlation function.
In the language of the O($n$) loop model, this correlation function
between two points separated by a  distance $r$  assumes the form
$g(r)=Z'/Z$, where $Z$ is defined by Eq.~(\ref{Zloop}) and $Z'$ by a
similar sum on graphs ${\cal G'}$ which include, apart from a number
of closed loops, also a single segment which connects the two
correlated points. We take these points far apart in the length direction
of the cylinder. For the calculation of $Z'$ we use the `odd' transfer
matrix built on connectivities that describe, in addition to a number of
pairwise connected points, also a single point representing the 
position of the unpaired loop segment. The `even' transfer matrix
uses only connectivities representing pairwise coupled points.

Let the largest eigenvalue of the even transfer matrix for a system of
size $L$ be $\Lambda^{(0)}_{L}$, and that of the corresponding odd
transfer matrix $\Lambda^{(1)}_{L}$. These eigenvalues determine the
magnetic correlation length $\xi_h(L)$ along a cylinder of size $L$:
\begin{equation}
\xi^{-1}_h(L)=(2/\sqrt{3})\ln(\Lambda^{(0)}_L/\Lambda^{(1)}_L)
\end{equation}
where the geometric factor $2/\sqrt{3}$ is the ratio between the
small diameter of an elementary face $(\sqrt{3})$ and the length added 
to the cylinder by each transfer matrix multiplication (3/2).
The present calculations of the relevant eigenvalues of the even and odd
transfer matrices are restricted to translationally invariant eigenstates.

The magnetic correlation length could thus be calculated numerically 
for finite sizes $L$ up to 10. Its dependence on the vacancy weight,
the bond weight, and $L$ allows the exploration of the phase diagram.
Suppose that the parameters of the model are near a renormalization
fixed point, at a distance $t$ along some temperature-like field direction.
The finite-size scaling behavior of the magnetic correlation length is
\begin{equation}
\xi_h^{-1}(t,L)=L^{-1}\tilde{\xi}_h^{-1}(tL^{Y_t})
\end{equation}
For sufficiently small $t$ we may expand the scaling function
$\tilde{\xi}_h$:
\begin{equation}
\xi_h^{-1}(t,L)=L^{-1}(\tilde{\xi}_h^{-1}(0)+ a tL^{Y_t} +\cdots )
\end{equation}
where $a$ is an unknown constant. For conformally invariant models,
$\tilde{\xi}_h^{-1}(0)= 2 \pi X_h$ where $X_h$ is the magnetic scaling 
dimension of the pertinent fixed point \cite{Cardy-xi,3Cardy}. This leads
us to define the scaled gap $X_h(t,L)\equiv L/(2 \pi \xi_h(t,L))$
which scales as
\begin{equation}
X_h(t,L)=X_h+ a' tL^{Y_t} +\cdots 
\end{equation}
where further contributions may be due to irrelevant scaling fields, 
or a second relevant temperature-like field if the system is close to
a tricritical point.
In actual calculations, we obtain the scaled gaps as a function of $y$ and
$w$, and their interpretation in terms of scaling fields involves one
relevant field in the case of an ordinary critical point, and two in the
case of a tricritical point. In order to get rid of the most relevant
field we may e.g. fix $y$ to some value and solve for $w$ in
\begin{equation}
X_h(y,w,L)=X_h(y,w, L+1)
\label{findcp}
\end{equation}
The solutions, which are denoted $w(y,L)$, are expected to approach the
actual critical point as $w(y,L)=w(y,\infty)+ c L^{Y_{t,2}-Y_t}$ where
$Y_t$ is the renormalization exponent of the most relevant temperature
field and $Y_{t,2}$ the second largest temperature-like exponent.
Thus, by solving Eq.~(\ref{findcp}) for a range of $L$ values we obtain a
series of estimates of the critical point $w(y,L)$. In the case of an
ordinary critical point, the irrelevance of other temperature-like fields
($Y_{t,2}<0$) implies that the values $X_h(y,w,L)$ taken at the solutions of
Eq.~(\ref{findcp}) for large $L$ converge to $X_h$ as
\begin{equation}
X_h(y,w(y,L),L)-X_h \propto L^{Y_{t,2}}
\label{findXh}
\end{equation}
By taking $w(y,L)$ we have effectively removed the most relevant field.
This formula still holds when $y$ has a value such that $y,w(y,\infty)$
is in the vicinity of a tricritical point, but then the second temperature
exponent is relevant: $Y_{t,2}>0$. In that case,
the values $X_h(y,w,L)$ taken at the solutions of Eq.~(\ref{findcp})
appear to {\em diverge} from the fixed-point value of $X_h$, i.e.
the tricritical magnetic exponent.

These considerations allow us to locate a tricritical point, if present.
Convergent behavior of $X_h(y,w(y,L),L)$ corresponds with ordinary
critical points, or with first-order transitions to which similar
arguments apply. Divergent behaviour, i.e. the $X_h(y,w(y,L),L)$
are moving away from the tricritical value of $X_h$ when $L$ increases,
reveals a tricritical point. Only when $y$ is set precisely at its
tricritical value, $X_h(y,w(y,L),L)$ will converge.

\section{Numerical results}

A superficial exploration of the O(0) model Eq.~(\ref{Zloop}) by means
of transfer-matrix calculations using small system sizes indicated that
it has a phase diagram similar to that found in Ref.~\cite{BBN}: the
low-temperature and high-temperature phases are separated by a line
of phase transitions which includes a range of ordinary O(0) critical
points, separated from a range of first-order transitions by an
exactly known \cite{DS,3BN} theta point. Furthermore, this
preliminary work indicated that the phase behavior does not change
in a qualitative sense in the range $0 \leq n \leq 1$.
Encouraged by this outcome we performed more extensive analyses for
$n=0$, $\half$, and 1. We obtained solutions of Eq.~(\ref{findcp}) for
fixed values $y=0$, 0.1, $\cdots$, 0.8 using system sizes up to $L=10$.

The results for $X_h(y,w(y,L),L)$ are shown in Figs. 1-3. For small
$y$ the data tend to converge to the known magnetic
dimensions \cite{3Nhon} for the O($n$)
critical points, indicated by dashed lines. For large $y$ values 
the data tend to converge to a much lower value of $X_h$ which does
not only reflect the expected behavior near a discontinuity fixed 
point \cite{NN} but also the algebraic nature of the dense O($n$)
phase. For intermediate $y$ we observe the behavior formulated above
for the vicinity of a
tricritical point: the data indicate that there exists a special value
of $y$ for which the $X_h(y,w(y,L),L)$ converge to a constant level
which is different from the large- and small $y$ asymptotes.

For $n=0$ (see Fig. 1) the tricritical point is
located at $y=\half$; the numerical data coincide with the exact
tricritical value $X_h=0$ for all $L$. For $n=\half$ we do not know
the exact location of the tricritical point and its associated
magnetic exponent, but from the data in Fig. 2 we conclude that it is
located near $y=0.54$, $w=1.058$, and that $X_h\approx 0.036$ (dotted
line).  For $n=1$ we estimate the tricritical point at $y=0.58$,
$w=1.115$. The finite-size data (see Fig. 3) are in a good agreement
with the exactly known tricritical value $X_h=3/40$ which is indicated
by a dotted line. As expected, for increasing $L$ the data are moving
away from that line.

The solutions $w(y,L)$  seem to converge well with increasing $L$.
The phase transitions thus found are shown as data points in the
phase diagrams of Figs. 4-6,  which apply to $n=0$, $\half$ and 1
respectively. The estimated errors are much smaller than the 
symbol sizes.

For $y=0$ our numerical procedures accurately reproduce the exact
critical points given in Ref.~\cite{3Nhon}: $w=0.54120\cdots$ for $n=0$,
$w=0.55687\cdots$ for $n=\half$, and $w=0.57735\cdots$ for $n=1$. These
exact critical points are shown as black circles.  For $n=0$ (Fig. 4)
the exactly known tricritical point (black square) coincides with
the data point for $y=\half$. For $n=\half$ and $n=1$ the estimated
tricritical points are shown as black squares. The data points in
each of Figs. 4-6 are connected by a line of phase transitions that 
divides into a continuous part (dashed line) and a first-order part
(full curve). These curves do not represent any further data but serve
only as a guide to the eye.

\section{Conclusion and outlook}

The numerical analysis of the O($n$) model with vacancies demonstrates
that one can interpolate between the known tricritical points at $n=0$
and $n=1$. These two points are thus not isolated cases, but special
cases of generic O($n$)
tricriticality. In spite of the apparent difficulty to find a class
of tricritical points analytically, the existence of such a class is
obvious. Forthcoming numerical analysis will show whether this family
of tricritical points covers the whole range $-2 \leq n \leq 2$ for
which analytical results for the ordinary O($n$) critical point are
known. At present we see no obvious reason for a negative answer. 

The tricritical exponents remain to be determined; our present work   
contributes only $X_h\approx 0.036$ as a new result for the O($\half$) 
model. This determines the decay of the magnetic correlation function 
with distance $r$ according to $r^{-\eta}$ with $\eta=2 X_h$. 
As demonstrated e.g. in Ref.~\cite{3BN}, several other
exponents can be determined such as the temperature dimension $X_t$
and magnetic exponents associated with different types of interfaces.
A further characterization of the family of O($n$) tricritical 
points will also require a determination of the conformal  anomaly.
While this is possible in principle \cite{BCN}, it remains to be seen
whether the inaccuracies involved in the determination of the
tricritical points will allow the emergence of a clear picture.
The present calculations are still rather limited in magnitude; only
a few hours of CPU time on a few Silicon Graphics workstations were
used, which we found to be sufficient for the present purposes. Thus,
the use of larger computer systems will open the possibility to explore
significantly larger system sizes $L$, and thus help to conduct such
more detailed investigations.

We conclude with a few remarks on the relation between O($n$)
tricriticality and the percolation of the vacancies. For $n=0$ the
vanishing loop weight enforces a perfect loop vacuum on the left hand
side of the line of phase transitions (Fig. 4). Thus, the site
percolation on the triangular lattice applies: the vacancy
percolation line lies at $y=\half$ and ends at the tricritical point.
Once the density of the vacancies has increased above the percolation
threshold, there is no way in which the loop configurations can
reach criticality, and the transition becomes discontinuous.
We thus expect that, also for $n\neq0$, the vacancy percolation
line connects to the tricritical point. However, for $n>0$ the loop
vacuum is no longer perfect so that the percolation threshold moves to
$y>\half$, in accordance with the estimated tricritical values of $y$.

\vspace{5mm}
{\em Acknowledgements}: 
We are indebted to Murray Batchelor, whose contributions to 
Ref.~\cite{BBN} were also valuable for the present work. This research 
is supported in part by the FOM
(`Stichting voor Fundamenteel Onderzoek der Materie') which is
financially supported by the NWO (`Nederlandse Organisatie voor
Wetenschappelijk Onderzoek').

\newpage
\begin{figure}
\epsfxsize=120mm
\epsffile{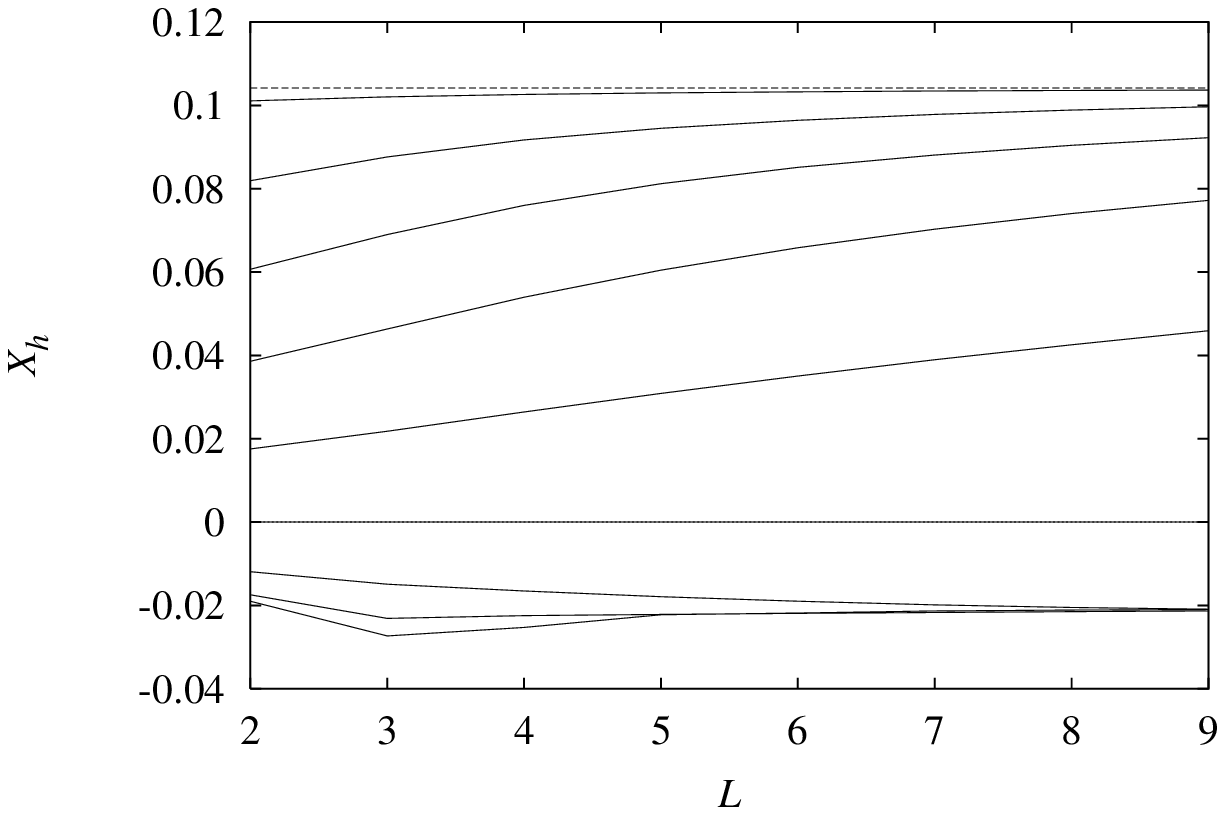}
\caption{Effective magnetic scaling dimension $X_h$ of the O(0) model
versus  system size $L$, for various values of the vacancy weight $y$.
The data points (not shown) are connected by lines. Counting from above,
these lines correspond with $y=0$, 0.1, 0.2, $\cdots$, 0.8. For small
$y$ the data behave in accordance with the expected magnetic dimension
$X_h=5/48=0.10416\cdots$ of the O($0$) critical point (dashed line).
The data for $y=0.5$ agree with the exactly known result for theta
point of this model.}
\label{xh1}
\end{figure}
\vspace{10mm}
\begin{figure}
\epsfxsize=120mm
\epsffile{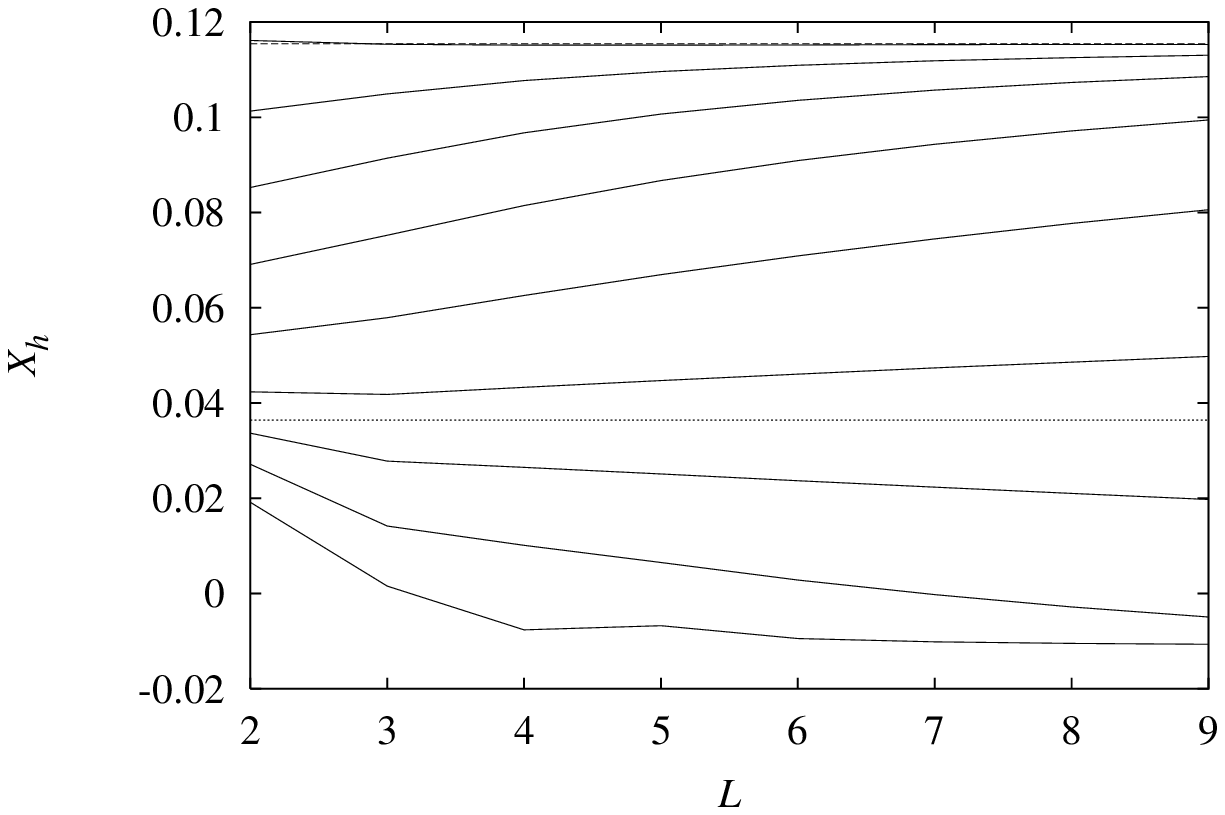}
\caption{Effective magnetic scaling dimension $X_h$ of the O($\half$) model
versus  system size $L$, for various values of the vacancy weight $y$.
The data points (not shown) are connected by lines. Counting from above,
these lines correspond with $y=0$, 0.1, 0.2, $\cdots$, 0.8. For small
$y$ the data behave in accordance with the expected magnetic dimension
$X_h=0.11544\cdots$ of the O($\half$) critical point (dashed line).
The tricritical magnetic exponent is estimated as $X_h\approx 0.036$
(dotted line).}
\label{xh2}
\end{figure}
\vspace{10mm}
\begin{figure}
\epsfxsize=120mm
\epsffile{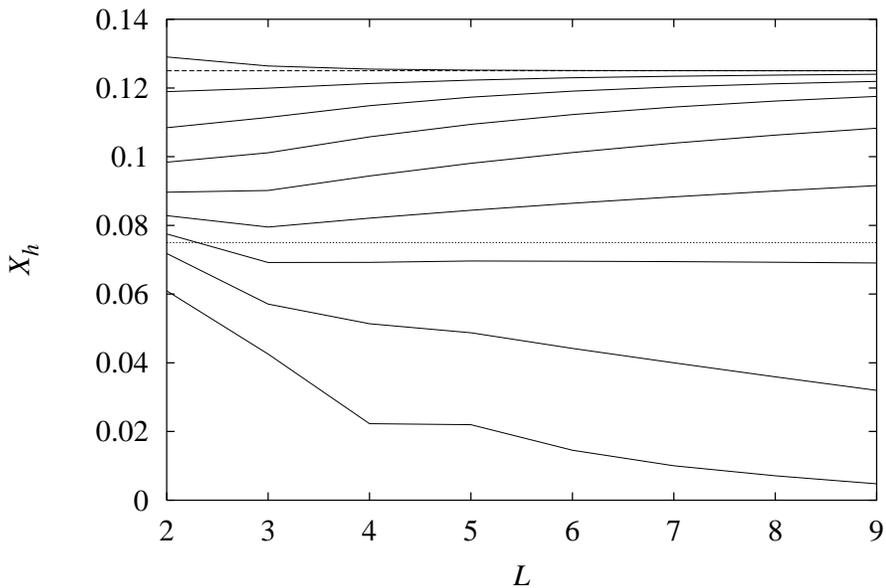}
\caption{Effective magnetic scaling dimension $X_h$ of the O(1) model
versus  system size $L$, for various values of the vacancy weight $y$.
The data points (not shown) are connected by lines. Counting from above,
these lines correspond with $y=0$, 0.1, 0.2, $\cdots$, 0.8. For small
$y$ the data behave in accordance with the expected magnetic dimension
$X_h=1/8$ of the Ising  critical point (dashed line). The exactly known
value of the tricritical Ising magnetic exponent $X_h=3/40$ is indicated
by a dotted line.}
\label{xh3}
\end{figure}
\vspace{10mm}
\begin{figure}
\epsfxsize=120mm
\epsffile{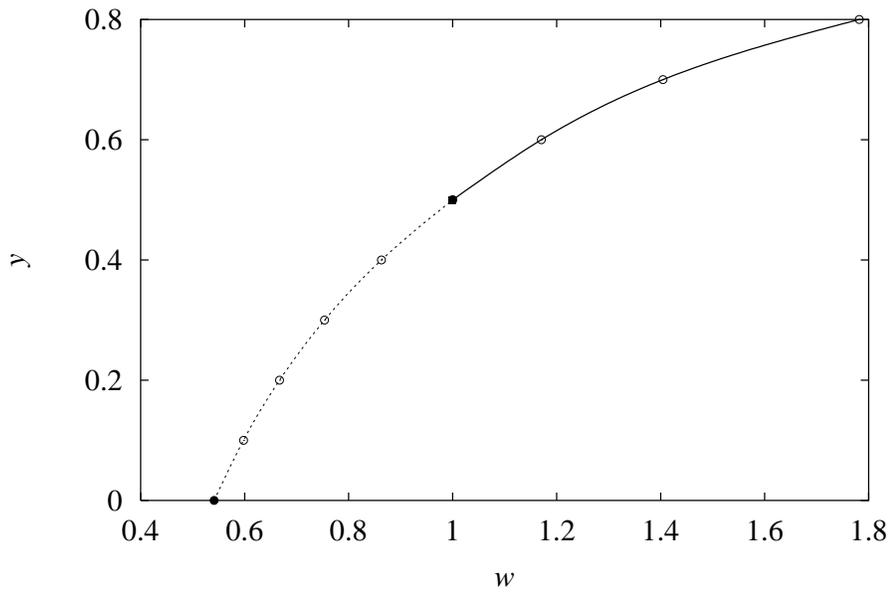}
\caption{Phase diagram of the O(0) model in the bond weight $w$ 
versus vacancy weight $y$ plane. The exactly known point at $y=0$ is
shown as a black circle, that at $y=\half$ (the theta point) as a black
square. The numerical data points are shown as open circles, connected
by a curve. The dashed part indicates critical points in the O(0)
universality class, the full part indicates first-order transitions. }
\label{tro1}
\end{figure}
\vspace{10mm}
\begin{figure}
\epsfxsize=120mm
\epsffile{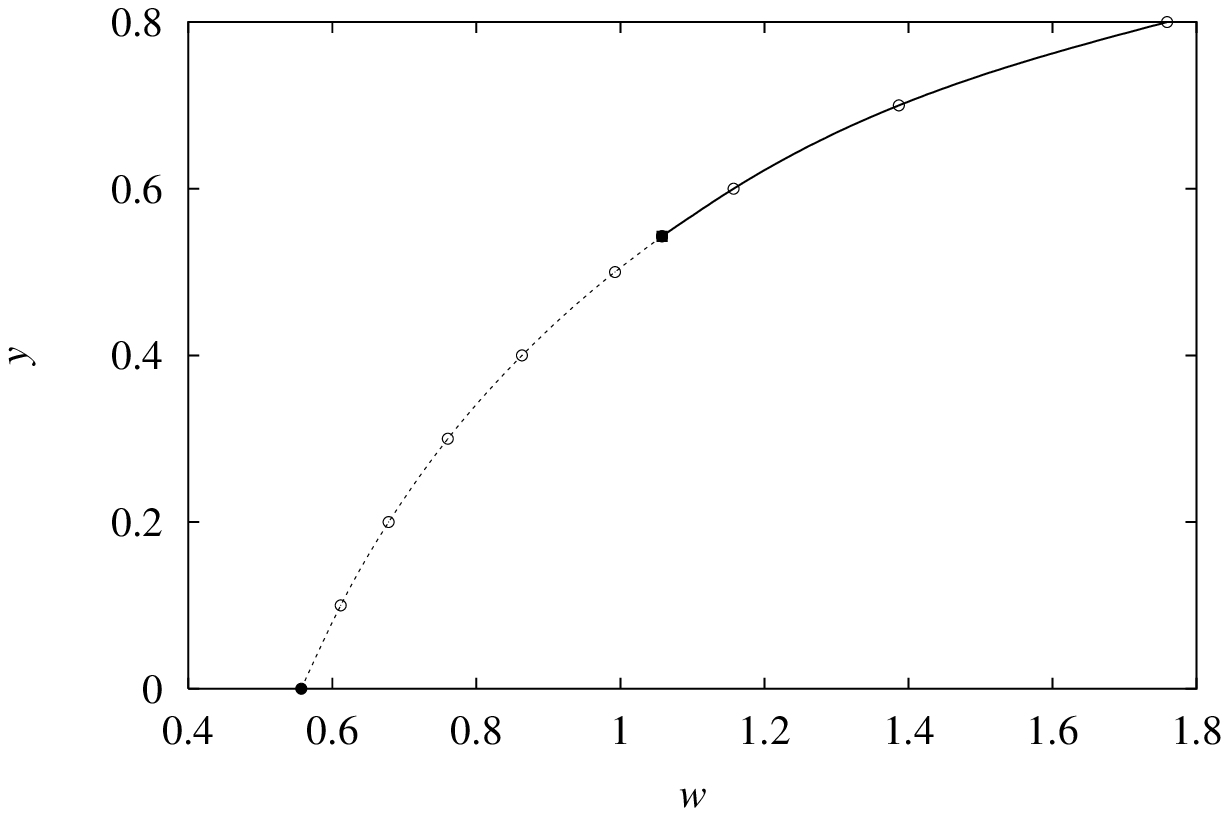}
\caption{Phase diagram of the O($\half$) model in the bond weight $w$ 
versus vacancy weight $y$ plane. The exactly known point at $y=0$ is
shown as a black circle.
The numerical data points are shown as open circles, connected
by a curve. The dashed part indicates critical points in the O($\half$)
universality class, the full part indicates first-order transitions. 
The estimated location of the tricritical point is shown as a black
square.}
\label{tro2}
\end{figure}
\vspace{10mm}
\begin{figure}
\epsfxsize=120mm
\epsffile{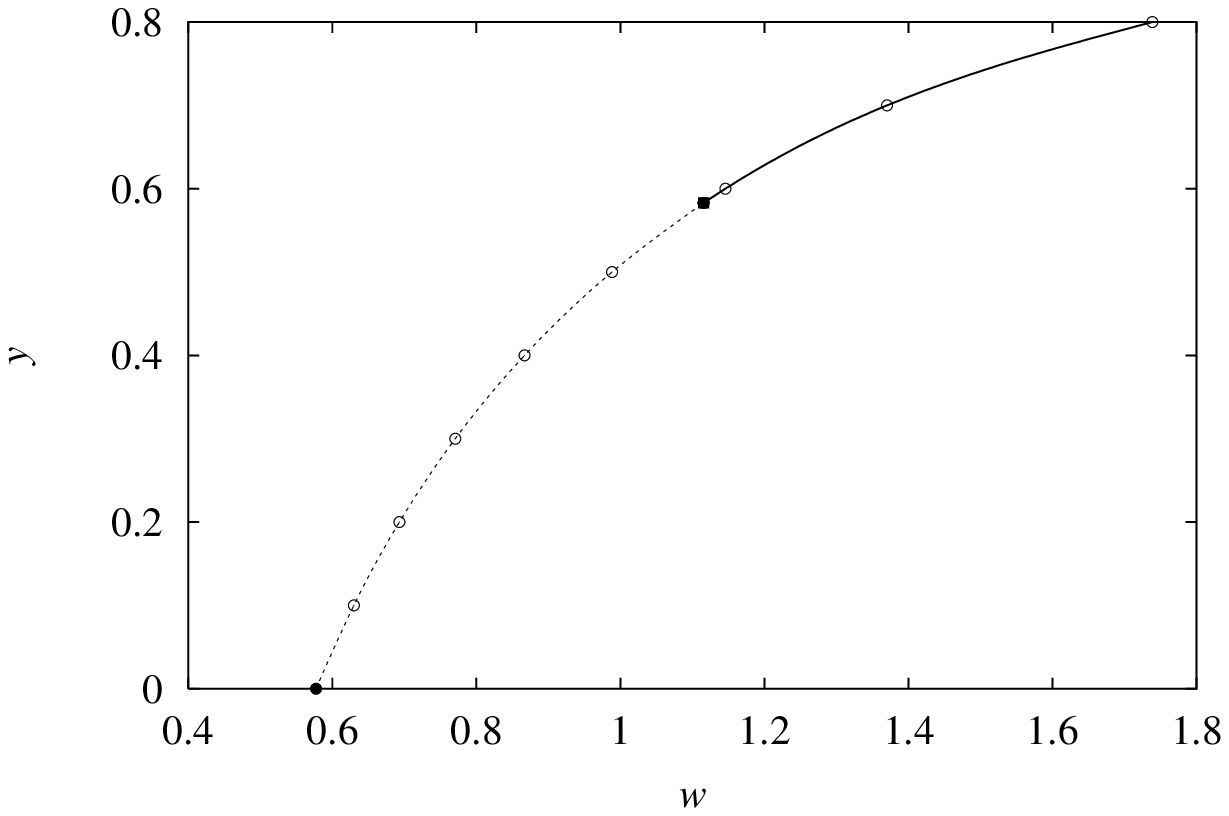}
\caption{Phase diagram of the O(1) model in the bond weight $w$
versus vacancy weight $y$ plane. The exactly known point at $y=0$ is
shown as a black circle.
The numerical data points are shown as open circles, connected
by a curve. The dashed part indicates critical points in the O(1)
universality class, the full part indicates first-order transitions. 
The estimated location of the tricritical point is shown as a black
square.}
\label{tro3}
\end{figure}
\end{document}